\documentclass[aps,prl,showpacs,twocolumn,superscriptaddress,nofootinbib]{revtex4-1}
\pdfoutput=1
\usepackage{graphicx}
\usepackage{color}      

\usepackage{bm}
\usepackage{amsmath}
\usepackage{times, txfonts}

\newcommand{\be}{\begin{equation}}
\newcommand{\ee}{\end{equation}}
\newcommand{\bea}{\begin{eqnarray}}
\newcommand{\eea}{\end{eqnarray}}

\def\a{\alpha}
\def\b{\beta}

\def\d{\delta}
\def\g{\gamma}

\def\ran{\rangle}
\def\lan{\langle}

\begin{document}

\title{Isotopic chains around oxygen from evolved chiral two- and three-nucleon interactions}

\author{A. Cipollone}
\affiliation{Department of Physics, University of Surrey, Guildford GU2 7XH, UK}

\author{C. Barbieri}
\email{C.Barbieri@surrey.ac.uk}
\affiliation{Department of Physics, University of Surrey, Guildford GU2 7XH, UK}

\author{P. Navr\'atil}
\affiliation{TRIUMF, 4004 Westbrook Mall, Vancouver, BC, V6T 2A3, Canada}

\date{\today}

\begin{abstract}
 We extend the formalism of self-consistent Green's function theory to include three-body interactions and apply it to isotopic chains around oxygen for the first time.
 The third-order algebraic diagrammatic construction  [ADC(3)] equations for two-body Hamiltonians can be exploited upon defining system-dependent one- and two-body interactions coming from the three-body force, and correspondingly dropping interaction-reducible diagrams. 
 The Koltun sum rule for the total binding energy acquires a correction due to the added three-body interaction. 
 This formalism is then applied to study chiral two-t and three-nucleon forces evolved to low momentum cutoffs. 
 The binding energies of nitrogen, oxygen and fluorine isotopes are reproduced with good accuracy and demonstrate of the predictive power of this approach. 
 Leading order three-nucleon forces consistently bring results close to the experiment for all neutron rich isotopes considered and reproduce the correct driplines for oxygen and nitrogen.
 The formalism introduced also allows to calculate  form factors for nucleon transfer on doubly magic systems.
 \end{abstract}

\maketitle

{\it Introduction}. -  The ultimate goal of ab-initio nuclear theory is to achieve accurate predictions of nuclear properties that are consistent with the underlying theory of QCD.  Advancing on this problem is presently of primary importance in the mid mass region of the nuclear chart, in response to significant advances in the discovery of new nuclides at radioactive isotope facilities~\cite{Thoe:2011}. Moreover, parameter free predictions would help reducing uncertainties in our knowledge of those dripline isotopes that are currently beyond experimental reach~\cite{Erler:2012}. 
 It has now become clear that accurate predictions require the explicit inclusion of multi-nucleon forces~\cite{Otsuka2010,Hagen12b,Holt2013}. For the oxygen chain, it has been shown that the Fujita-Miyazawa  three-nucleon force (3NF) is responsible for explaining the anomalous dripline at $^{24}$O~\cite{Otsuka2010}.  Ref.~\cite{Hagen12a} confirmed this result by considering approximated chiral 3NFs  at leading order (N$^2$LO). However, no investigation of 3NF's effects on neighboring isotopic chains has been made to date.
 In this Letter, we find that a correct inclusion of N$^2$LO 3NFs consistently reproduces the observed binding energies and that 3NFs similarly affect the behaviour near the driplines for other  isotopes as well. 
 
Concerning the calculation of mid mass nuclei, breakthroughs  were possible over the last decade due to the introduction of many-body methods that scale gently with increasing particle number. Self consistent Green's function theory (SCGF) \cite{Barbieri:2009nx,Barbieri:2009ej}, coupled cluster (CC) \cite{Hagen:2010gd,Hagen12a,Hagen12b} and in-medium similarity renormalization group (IM-SRG) \cite{Tsukiyama11,Hergert13a} have been employed in ab-initio calculations of doubly closed shell nuclei with masses up to A$\sim$60. For open-shells, semi-magic isotopes can be calculated by breaking particle conservation symmetry and  reformulating theories in terms of Hartree-Fock Bogolioubov reference states, as done in Gorkov theory \cite{VdSluys1993,Soma11,Idini2012,Soma13a} and in IM-SRG~\cite{Hergert13b}. Calculations based on IM-SRG have been performed for ground state energies. On the other hand, the state-of-the-art SCGF theory can also be extended to the Gorkov approach~\cite{Idini2012,VdSluys1993}  and it gives access to a wealth of nuclear structure information. This includes  the addition or removal of one or two nucleons to and from the calculated ground states~\cite{Barbieri:2006sq,Barbieri:2004Shh} and direct link to microscopic optical potentials~\cite{Charity2006,Barbieri:2005NAscatt}.

This Letter extends the scope of SCGF to include 3NFs in finite nuclei. We define density dependent one- and two-body interactions derived from the 3NF part of the Hamiltonian and work out the correction to the Koltun sum rule to obtain binding energies. This allows us to fully include   chiral 3NFs in the third-order algebraic diagrammatic construction [ADC(3)] equations commonly used in quantum chemistry applications~\cite{Schirmer1983,Barbieri2007}.
The method is applied to study chiral 3NFs in the oxygen, nitrogen and fluorine isotopic chains, as well as  the spectra of single neutron states in the {\em sd} shell.
This opens the possibility of probing modern realistic nuclear interactions on a wide range of experimental data, including excitation spectra,  the evolution of shell closures, and the position of driplines.

{\it Formalism}.
We employ Green's function (or propagator) theory and calculate the single particle propagator~\cite{FetWa:book},
\begin{align}
 g_{\alpha \beta}(\omega) ~=~& 
 \sum_n  \frac{ 
          \langle {\Psi^A_0}     \vert c_\alpha        \vert {\Psi^{A+1}_n} \rangle
          \langle {\Psi^{A+1}_n} \vert c^{\dag}_\beta  \vert {\Psi^A_0} \rangle
              }{\omega - \varepsilon_n^{A+1} + i \eta }  ~+~\nonumber\\
 ~+~ &\sum_k \frac{
          \langle {\Psi^A_0}     \vert c^{\dag}_\beta  \vert {\Psi^{A-1}_k} \rangle
          \langle {\Psi^{A-1}_k} \vert c_\alpha        \vert {\Psi^A_0} \rangle
             }{\omega - \varepsilon_k^{A-1} - i \eta } \; ,
\label{eq:g1}
\end{align}
where greek indices $\alpha$,\,$\beta$,..., label a complete orthonormal basis set
and $\varepsilon_n^{A+1}\equiv(E^{A+1}_n - E^A_0)$ and $\varepsilon_k^{A-1}\equiv(E^A_0 - E^{A-1}_k)$
are the nucleon addition and separation energies, respectively.
In Eq.~(\ref{eq:g1}),  $\vert\Psi^{A+1}_n\rangle$, $\vert\Psi^{A-1}_k\rangle$ are the eigenstates and $E^{A+1}_n$, $E^{A-1}_k$ are the eigenenergies of the ($A\pm1$)-nucleon system.  Hence, $g_{\alpha \beta}(\omega)$  describes the exact propagation of a single nucleon and hole excitations through the system.
From Eq.~(\ref{eq:g1}) we also extract the one-body reduced density matrix,
\begin{align}
 \rho_{\a\b} ~=~&\lan{\Psi^A_0} | c^{\dag}_\b c_\a |{\Psi^A_0} \ran ~= ~\int_{C\uparrow} \frac{d\omega}{2 \pi i} \; g_{\a\b}(\omega)  \; ,
\label{eq:rho} 
\end{align}
where the integration contour $C\!\uparrow$ is taken on the upper half of the imaginary plane.


We start our calculations with the intrinsic Hamiltonian  $H(A)= H - T_{c.m.}(A) = U(A) + V(A) + W$ in which the kinetic energy of the center of mass (c.o.m.) has been subtracted and we put in evidence the dependence on the number of nucleons~$A$.  The terms $U$, $V$ and $W$ collect all the one-, two- and three-nucleon contributions, respectively.
Based on this, we define system dependent one- and two-body effective interactions obtained by contraction with the correlated density matrix, Eq. (\ref{eq:g1}),
\begin{align}
\tilde{U}_{\alpha \beta}  & = U_{\alpha \beta}   +  V_{\alpha \gamma, \, \beta \delta} \,  \rho_{\delta\gamma}
                                  +  \frac{1}{2}  W_{\alpha \gamma \delta, \, \beta \mu \nu} \, \rho_{\mu \gamma} \, \rho_{\nu \delta}  \; ,
\label{eq:Ueff} \\
\tilde{V}_{\alpha \beta, \gamma \delta} & =  \qquad \quad V_{\alpha \beta, \gamma \delta}  \quad + \quad 
                W_{\alpha \beta \mu, \gamma \delta \nu} \, \rho_{\nu \mu} \; .
\label{eq:Veff}
\end{align}
All matrix elements are properly antisymmetrized and summation over repeated indices are implied here and in the following.
The resulting Hamiltonian, $\widetilde{H}=\widetilde{U}+\widetilde{V}+W$, can be proved to lead to the same Green function (\ref{eq:g1}) as the original Hamiltonian with the caveat that only {\em interaction-irreducible} terms are retained in the diagrammatic expansion~\cite{CarboneIP}\footnote{A diagram is said to be {\em interaction-reducible} if it can be factorized in two lower-order diagrams by cutting an interaction vertex or, equivalently, if it is connected and there exists a group of lines (interacting or not) that leave an interaction vertex and eventually all return to it.}.
Equations (\ref{eq:Ueff}) and (\ref{eq:Veff}) generalize the idea of normal ordering of the Hamiltonian to fully correlated densities.  In this work we keep only the $\widetilde{U}$ and $\widetilde{V}$ terms and discard diagrams with explicit interaction-irreducible 3NFs. The error associated with this truncation has been seen to be negligible in Refs.~\cite{Hagen2007,Binder13a}~\footnote{This would be the analogous to the NO2B approximation of Ref.~\cite{Binder13a} but, here, not tight to the choice of any reference state.}.
 The single particle propagator $g_{\alpha\beta}(\omega)$ can then be calculated by exploiting the effective one- and two- body interactions with the already available two-body formalisms. 

We first solve the spherical Hartree-Fock (HF) equations for the full Hamiltonian within the given model space. The resulting propagator, $g^{HF}_{\alpha\b}(\omega)$, is then used as a reference state to calculate the energy-dependent part of the self-energy. 
We employ ADC(3) method~\cite{Schirmer1983,Barbieri2007} and write the self-energy as
\begin{align}
  \Sigma_{\a\b}^{\star}(\omega)=&\Sigma_{\a\b}^{\infty}+\Sigma_{\a\b}'(\omega)
\label{eq:Self}  \\
= &\tilde{U}_{\a\b}+C_{\a n}\left[\frac{1}{\omega-{\bm M}}\right]_{n \, n'}C^{\dag}_{n' \b}+D_{\a k}\left[\frac{1}{\omega-{\bm N}}\right]_{k \, k'}D^{\dag}_{k' \b} \; ,
\nonumber
\end{align}
where {\bf M}~({\bf N}) are interaction matrices in the {\em 2p1h}~({\em 2h1p}) space and {\bf C}~({\bf D})  are the corresponding coupling strengths to the single particle states. In the ADC(3), these matrices are constructed to guarantee that all diagrams up to third order are included in Eq.(\ref{eq:Self}). In general, the the ADC($n$) approach defines a hierarchy of truncation schemes of Eq.~(\ref{eq:Self}) for increasing order $n$ that guides systematic improvements of the method. The correlated propagator $g_{\alpha \beta}(\omega)$ is finally obtained by solving the Dyson equation, 
\begin{equation}\label{eq:Dy}
g_{\alpha\beta}(\omega)=g^{HF}_{\alpha\b}(\omega)+g^{HF}_{\alpha\g}(\omega)\Sigma_{\g\d}^{\star}(\omega)g_{\d\b}(\omega) \, ,
\end{equation}
which is diagonalized using a Lanczos algorithm as explained extensively in~\cite{Schirmer1989,Soma13b}.
Note that  we employ  the {\em sc0} approximation of Refs.~\cite{Soma13a,Soma13b} where only the $\Sigma_{\a\b}'(\omega)$ contribution of Eq.~(\ref{eq:Self}) depends on the reference states $g^{HF}_{\alpha\b}(\omega)$. This implies the iterative solutions of Eq.~(\ref{eq:Dy}) to evaluate $\Sigma_{\a\b}^{\infty}$=$\tilde{U}_{\a\b}-U^{HF}_{\a\b}$ in terms of the final correlated density matrix, Eq.~(\ref{eq:Ueff}).

In the presence of 3NFs, the ground state energy can still be inferred from the Koltun sum rule (SR) that now acquires a correction:
\begin{equation}
\label{eq:koltun}
E^A_0 = \sum_{\alpha \, \beta} \frac{1}{4 \pi i} \int_{C \uparrow} d \omega  \; \;
\left[ U_{\alpha \beta}  +  \omega \delta_{\alpha \beta} \right] \, g_{\beta \alpha}(\omega)   
~ - ~ \frac{1}{2}\langle {\Psi^A_0}     \vert W     \vert {\Psi^A_0} \rangle  \; .
\end{equation}
Eq.~(\ref{eq:koltun})--based on the exact propagator--is still an exact equation. However, it requires to evaluate the expectation value of the 3NF part of the Hamiltonian $\langle {\Psi^A_0}     \vert W     \vert {\Psi^A_0} \rangle$, with an accuracy comparable to the many-body approximation in use.
We calculate this correction at first order in W using fully correlated propagators,
\begin{equation}\label{eq:W1st}
\langle W^{3\rho}\rangle=\frac{1}{6} W_{\alpha\beta\gamma, \, \mu\nu\xi}\,\,\rho_{\mu\alpha}\,\,\rho_{\nu \beta}\,\,\rho_{\xi \gamma} \, ,
\end{equation}
that implicitly includes relevant higher order terms from standard many-body perturbation theory.
We found that it is  mandatory to use fully dressed propagators--the solution of  Eq. \eqref{eq:Dy}--but that this is also sufficient to account for all relevant contributions.  The next order correction is given by
\begin{equation}\label{eq:Wtda}
\langle W^{TDA}\rangle=\frac{1}{4} W_{\alpha\beta\gamma,\mu\nu\xi}\,\,\rho_{\mu\alpha}\,\,\Delta\Gamma_{\nu\xi,\beta\gamma} \, ,
\end{equation}
where $\Delta\Gamma$ is the two-body density matrix after subtraction of the zeroth-order contribution coming from two {\em fully correlated} but {\em non-interacting} nucleons, to avoid double counting with Eq.~(\ref{eq:W1st}). We estimated this using in Tamn-Dancoff approximation (TDA)~\cite{ring80a} and found its contribution to be small compared to our estimated errors, as discussed below.

The binding energy and spectra of neighboring even-odd isotopes are extracted from the poles of propagator \eqref{eq:g1}, however this requires a proper correction to account for the variation in the kinetic energy of the c.o.m. motion with changing~$A$. To extract the energy of a system with mass $A\pm1$, we recalculate $g_{\alpha \beta}(\omega)$ for the doubly closed subshell A-nucleon system but with a $\widetilde{H}(A\pm 1)$ corrected Hamiltonian. We then obtain:
\begin{align}
E^{A\pm1}  = \pm  \varepsilon^{A\pm1}_0[\widetilde{H}(A\pm1)] + E^A_0[\widetilde{H}(A\pm1)] \, ,
\label{eq:com}
\end{align}
where we made explicit the dependence on the c.o.m. correction for the Hamiltonian and  $E^A_0[\widetilde{H}(A\pm1)]$ is calculated using the corrected Koltun SR.

\begin{figure}[t]
\includegraphics[height=0.76\columnwidth,clip=true]{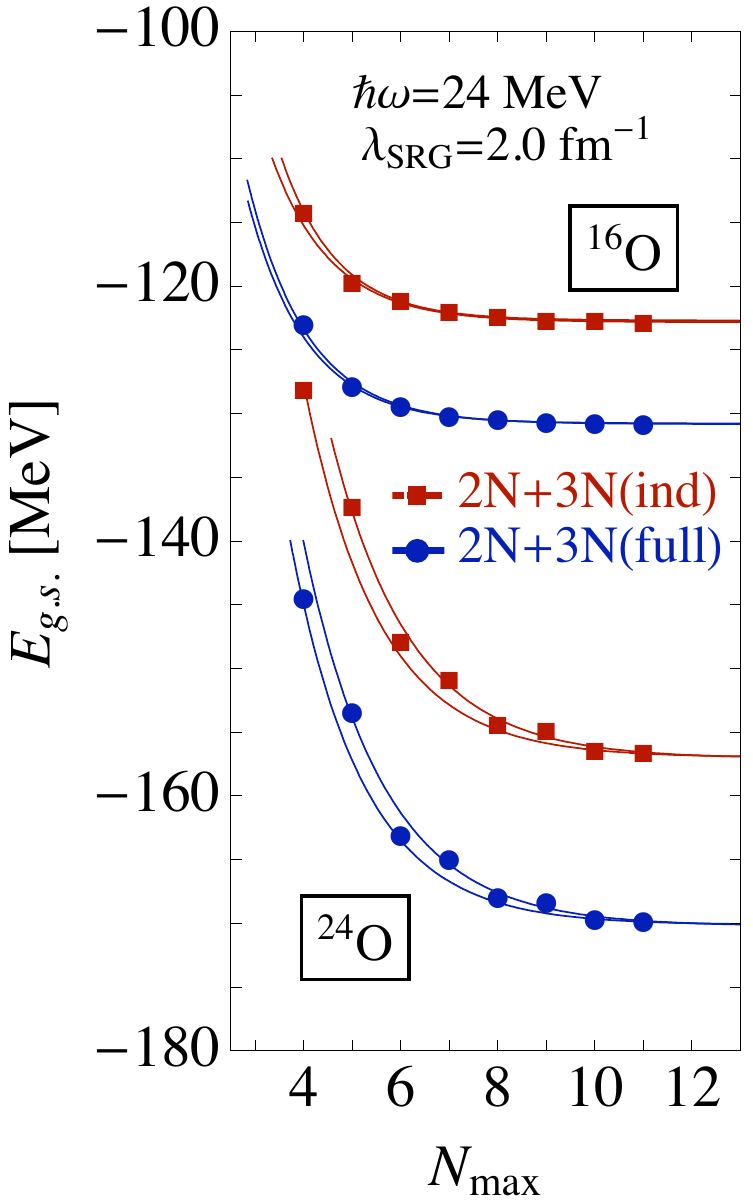}
\hspace{0.05in}
\includegraphics[height=0.76\columnwidth,clip=true]{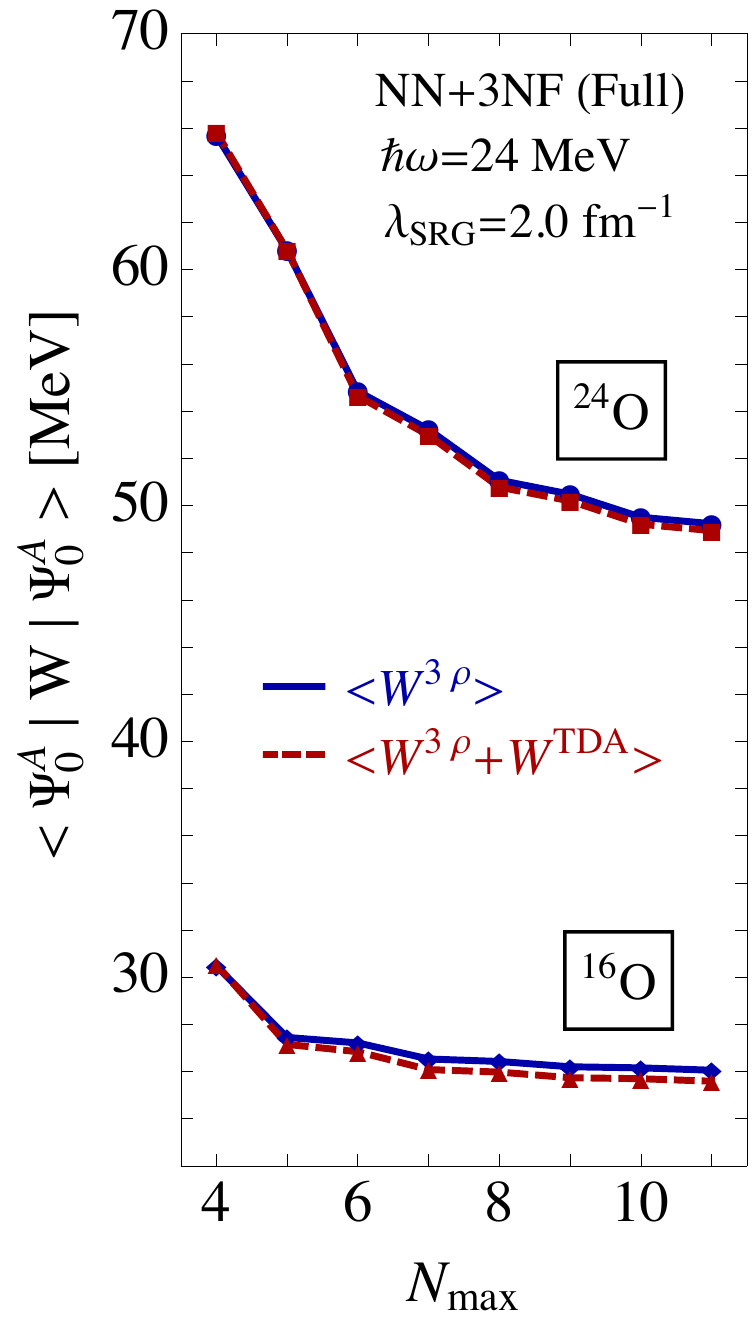}
\caption{(Color online)  
 Convergence of  $^{16}$O and $^{24}$O as a function of increasing size of the model space for $\hbar\omega$=24~MeV  and $\lambda_{SRG}$=2.0~fm$^{\rm -1}$. 
{\em Left:} Binding energies from the corrected Koltun sum rule, Eq.~(\ref{eq:koltun}), as obtained from the induced and full interactions.
{\em Right.}  Expectation value $\langle {\Psi^A_0}     \vert W     \vert {\Psi^A_0} \rangle$ obtained at first order only, Eq.~(\ref{eq:W1st}), (full lines) and with correction from two-body TDA ladders, Eq.~(\ref{eq:Wtda}) (dashed lines). 
}
\label{fig:conv}
\end{figure}

{\it Results}.
We perform calculations using chiral effective field theory (EFT) two-nucleon (2N) and 3NFs  evolved to low momentum scales by using free-space similarity renormalization group (SRG)~\cite{Jurgenson2009,Hebeler2012a}. The original 2N interaction is N$^3$LO with cutoff $\Lambda_{2N}$=500 MeV~\cite{Entem2003,Machleidt2011}. For the 3NF we use the N$^2$LO  interaction in a local form~\cite{Navratil2007} with a reduced cutoff of $\Lambda_{3N}$=400~MeV and low-energy constants $c_D$=-0.2 and $c_E$=0.098 refitted to reproduce the $^4$He binding energies, as discussed in Ref.~\cite{Roth2012prl}.
This choice of  $\Lambda_{3N}$ softens the contributions of two-pion 3NF terms, herby minimizing the impact of evolved 4NF.
The 3NF obtained by evolving only the original 2N-N$^3$LO Hamiltonian will be referred to as ``induced'' 3NF and it is independent of the pre-existing 3N-N$^2$LO interaction. Similarly, the ``full'' Hamiltonian is generated by evolving both initial 2N and 3NFs together. Since two-pion exchange diagrams that incorporate physics from the Fujita-Miyazawa 3NF appear at leading order, in the chiral 3N-N$^2$LO force, their effects are incorporated only in the full Hamiltonian.
Calculations were performed in model spaces up to 12 harmonic oscillator shells [$N_{max} \equiv$~max~$(2n+l)$ = 11], including all 2N matrix elements and limiting 3NF ones to configurations 
 with $N_1$+$N_2$+$N_3\leq N_{3NF,max}$=14. 

\begin{figure}[t]
\includegraphics[width=0.88\columnwidth,clip=true]{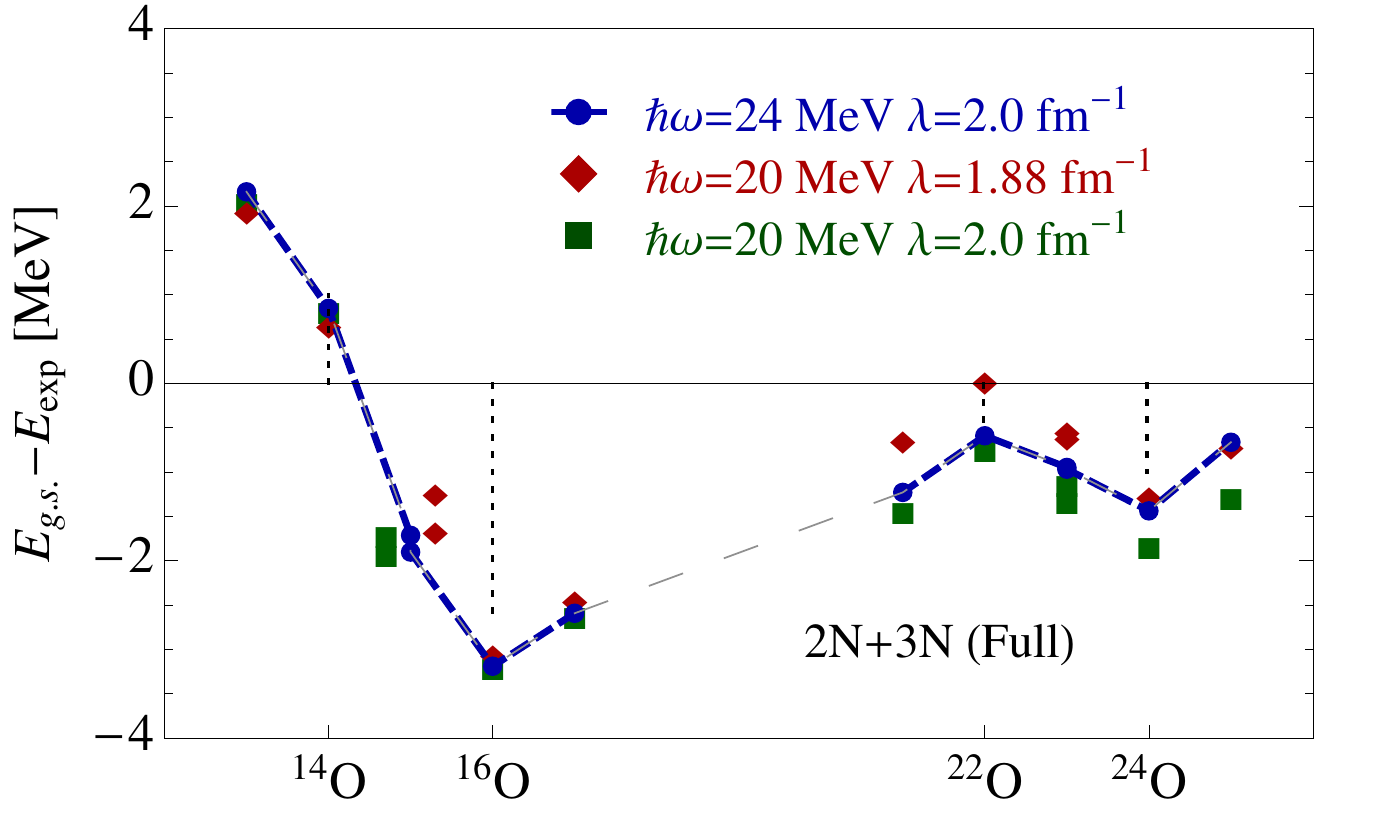}
\caption{(Color online) 
Differences between calculated and experimental ground state energies of oxygen isotopes for the full interaction with different values of $\hbar\omega$ and $\lambda_{SRG}$. 
Results for $^{15}$O  and $^{23}$O are obtained from two separate calculations, one for neutron addition and one for neutron removal (see text). Dots and diamonds for $^{23}$O are almost indistinguishable with the scale of this plot.
}
\label{fig:DeltaEvsA_O}
\end{figure}

Figure \ref{fig:conv} shows the convergence pattern of  total binding energies for $^{16}$O and $^{24}$O as a function of the model space size. The convergence is optimized by the choice of the chosen oscillator frequency which is close to the minimum of the $\hbar \omega$ dependence for the present interaction~
\cite{Hergert13a}. The staggering between adjacent results is due to the particular truncation of 3NF matrix elements and the alternate parities of harmonic oscillator shells. Separate exponential fits to the calculated  $^{24}$O energies, for $N_{max}$ either even or odd, differ by 100~keV  and are within 600~keV of the $N_{max}$=11 result.
Similarly, changing $\hbar\omega$ between 20 and 24~MeV we find about 450~keV variation in  our results (see Fig.~\ref{fig:DeltaEvsA_O}). Overall, these errors  amount to about 0.6\% of the total binding energy. 
The right panel of Fig.~\ref{fig:conv} demonstrates the similar convergence of the $\langle {\Psi^A_0}     \vert W     \vert {\Psi^A_0} \rangle$ expectation values. The contribution of $\langle W^{TDA}\rangle$, Eq.~(\ref{eq:Wtda}), is never bigger than 300~keV and represent a small correction to the Koltun SR. A proper study of the contributions to  $\langle W \rangle$ will be addressed in a forthcoming publication. For the present purposes, we sum the above uncertainties to make a conservative estimate for our convergence error of 1\%  for the calculated total binding energies.


The differences between calculated binding energies and the experiment, are demonstrated in Fig.~\ref{fig:DeltaEvsA_O} for different values of $\hbar\omega$ and $\lambda_{SRG}$. 
Refs.~\cite{Hergert13a,Binder13a} studied variations of $\lambda_{SRG}$ within larger intervals than the one considered here and found uncertainties of at most a few per cent.  This gives an estimate of the error committed by neglecting 4NF and higher terms. Our results with  $\lambda_{SRG}$ in the limited range 1.88-2.0~fm$^{\rm -1}$ do not exceed 0.5\% and this in agreement with Ref.~\cite{Binder13a}. 

From previous studies based on the ADC(3) method, we expect an accuracy of the many-body truncation scheme of about 1\%~\cite{Degroote2011,Barbieri2012}.
The extrapolated $^{16}$O ground state (Fig.~\ref{fig:conv}) is  over bound at -130.8(1)~MeV but in close agreement with the -130.5(1)~MeV obtained from IM-SRG~\cite{Hergert13a}, giving further  confirmation of the accuracy achieved by different many-body methods.
Note that the energies of $^{15}$O and $^{23}$O can be obtained in two different ways, from either neutron addition  or removal on neighboring sub-shell closures. Results in Fig.~\ref{fig:DeltaEvsA_O} differ by at most 400~keV, again within the estimated uncertainty of our many-body truncation scheme. The correction in Eq.~(\ref{eq:com}) is crucial to obtain this agreement. For  $\hbar\omega$=24~MeV and $\lambda_{SRG}$=2.0 fm$^{-1}$,  the discrepancy in $^{15}$O ($^{23}$O) is 1.65~MeV~(1.03±MeV) when  neglecting the changes in kinetic energy of the c.o.m. but it reduces to only  190~keV~(20~keV) when this is accounted for. 
This gives us confidence that a proper separation of the center of mass motion is being reached.
Fig.~\ref{fig:DeltaEvsA_O} also gives a first remarkable demonstration of the predictive power of chiral 2N+3N interactions: accounting for the precision of our many-body approach and dependence on $\lambda_{SRG}$ found in  Ref.~\cite{Binder13a},  we expect an accuracy of at least 5\% on binding energies. All calculated values agree with the experiment within this limits. Note that the interaction employed were only constrained by 2N and $^3$H/$^{4}$He data.

\begin{figure}[t]
\hspace*{0.1 cm}
\includegraphics[width=0.90\columnwidth,clip=true]{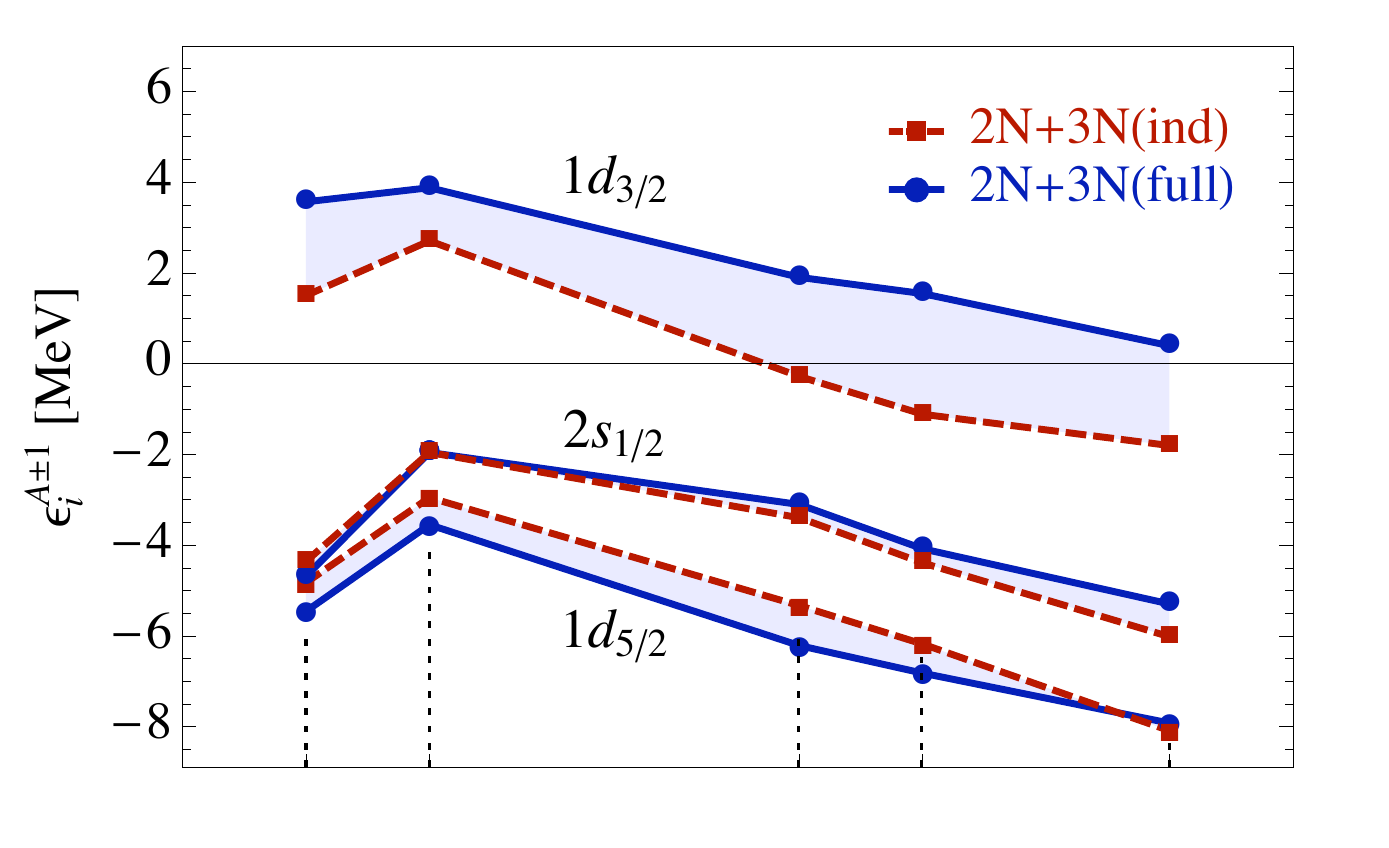}
\includegraphics[width=0.91\columnwidth,clip=true]{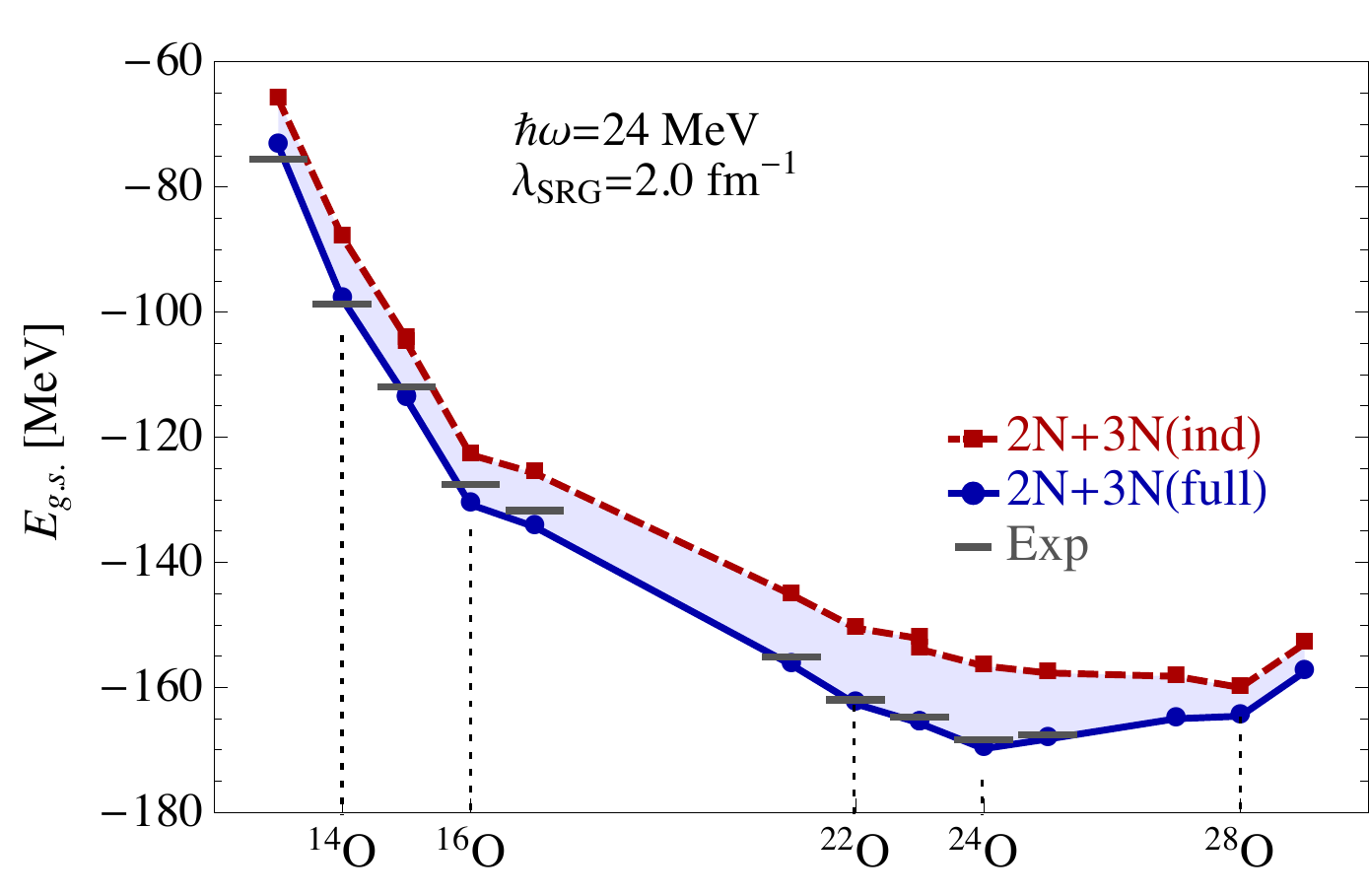}
\caption{(Color online)  
 {\em Top}. Evolution of single particle energies for neutron addition and removal around sub-shell closures of oxygen isotopes.
 {\em Bottom}. Binding energies obtained from the Koltun SR and the poles of propagator (\ref{eq:g1}), compared to experiment (bars)~\cite{AME2003,Jurado2007,Hoffman2008}.
 All points are  corrected for the kinetic energy of the c.o.m. motion.   For all lines, red squares (blue dots) refer to induced (full) 3NFs. 
}
\label{fig:EvsA_O}
\end{figure}

\begin{figure}[h]
\includegraphics[width=0.93\columnwidth,clip=true]{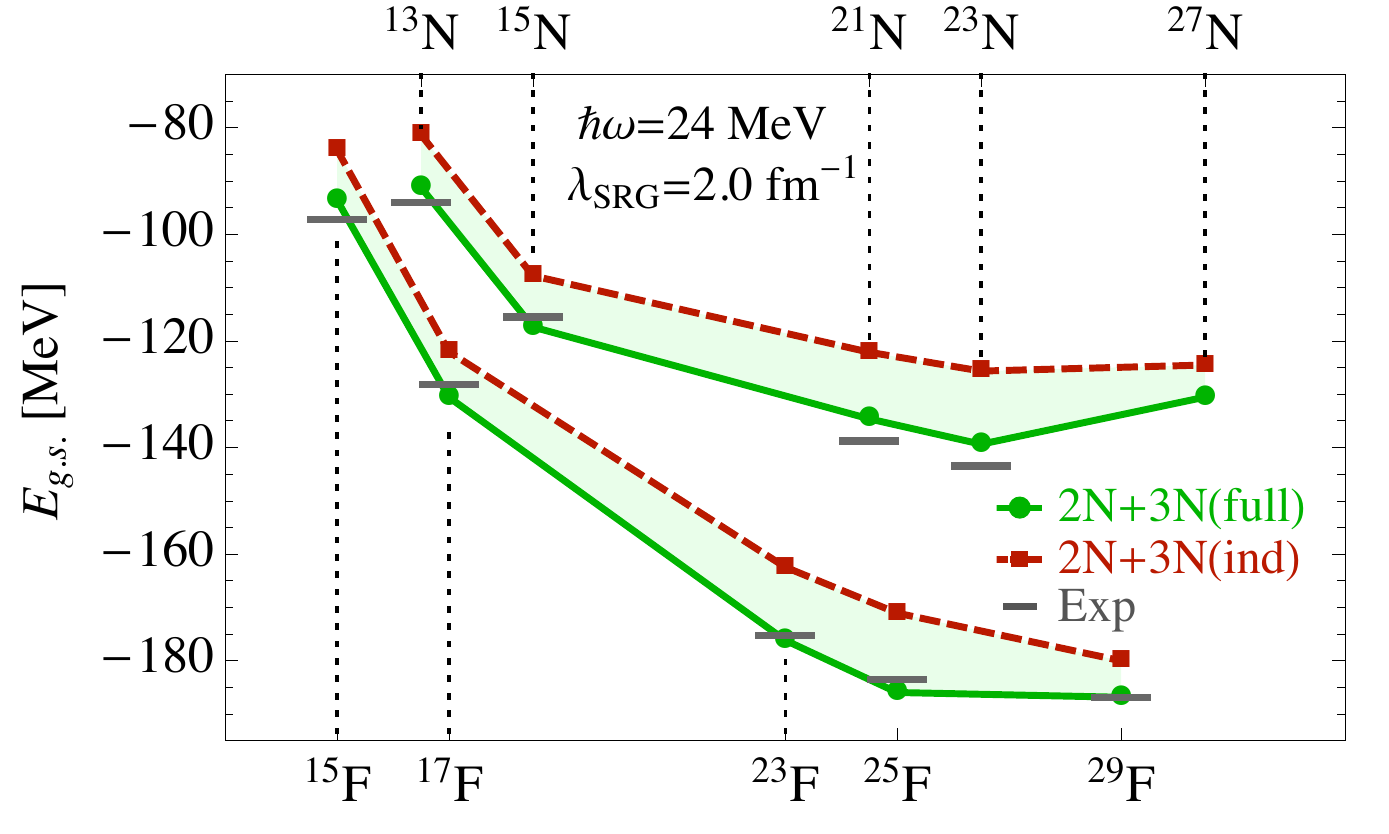}
\caption{(Color online)  
Binding energies of odd-even nitrogen and fluorine isotopes calculated for induced (red squares) and full (green dots) interactions.
Experimental data are from~\cite{AME2003,Jurado2007,Gaudefroy2012}.
}
\label{fig:EvsA_N_F}
\end{figure}

Figures~\ref{fig:EvsA_O} and~\ref{fig:EvsA_N_F} collects our results for the oxygen, nitrogen and fluorine isotopes calculated with $\hbar\omega$=24~MeV and $\lambda_{SRG}$=2.0 fm$^{\rm -1}$.
The top panel of Fig.~~\ref{fig:EvsA_O} shows the predicted evolution of neutron single particle spectrum (addition and separation energies) of oxygen isotopes in the {\em sd} shell. Induced 3NFs reproduce the overall trend but predict a bound $d_{3/2}$ when the shell is filled.  Adding pre-existing 3NFs---the full Hamiltonian---raises this orbit above the continuum also for the highest masses.
This gives a first principle confirmation of the repulsive effects of the two-pion exchange Fujita-Miyazawa interaction discussed in Ref.~\cite{Otsuka2010}. 
The consequences of this trends are demonstrated by the calculated ground state energies shown in the bottom panel: the induced Hamiltonian systematically under binds  the whole isotopic chain and erroneously places the  dripline at $^{28}$O due to the lack of repulsion in the $d_{3/2}$ orbit. Pre-existing 3NFs are substantial and increase with the mass number up to $^{24}$O, when the unbound $d_{3/2}$ orbit starts being filled.  At the same time, the $d_{5/2}$ quasiparticle states are lowered by about 1 MeV between $^{16}$O and $^{24}$O, which provides extra binding through the Koltun SR formula~(\ref{eq:koltun}) and corrects the slope of binding energies. As a result, the inclusion of N$^2$LO  3NFs  consistently brings calculations close to the experiment and reproduces the observed dripline at $^{24}$O~\cite{Hoffman2009,Kanungo2009,Janssens2009}.
Our calculations predict  $^{25}$O to be particle unbound by 1.54~MeV, larger thatn the experimental value of 770~keV~\cite{Hoffman2008} but within the estimated errors.  
The ground state resonance for $^{28}$O is suggested to be unbound by 5.2~MeV with respect to $^{24}$O. However this estimate is likely to be affected by the presence of the continuum which is important for this nucleus but neglected in the present work.

The same mechanism affects neighboring isotopic chains. This is demonstrated in Fig.~\ref{fig:EvsA_N_F} for the semi-magic odd-even isotopes of nitrogen and fluorine. Induced 3NF forces consistently  under bind these isotopes and even predict a $^{27}$N close in energy to $^{23}$N. This is fully corrected by full 3NFs that strongly binds $^{23}$N with respect to $^{27}$N, in accordance with the experimentally observed dripline.  The repulsive effects of filling the $d_{3/2}$ is also observed in $^{29}F$. However, the inclusion of an extra proton provides enough extra binding to keep this isotope bound by about 890~keV (calculated) with respect to $^{25}F$,  closer to the experimental value of 3.55 MeV~\cite{Gaudefroy2012}. The induced interaction alone would overestimate this binding and N$^2$LO  3NFs are fundamental in achieving the correct balancing between the attraction generated by the extra proton and the repulsion due to the filling of the neutron {\em sd}  shell.

In conclusion, we have considered the extension of the SCGF method to include three-body Hamiltonians. By properly defining system dependent effective one- and two-body interactions that include the relevant contribution from 3NFs, calculations can be performed with formalisms already existing for two-body Hamiltonians. 
This approach, however, goes beyond  usual truncations based on normal ordering of the Hamiltonian and employs  fully correlated densities instead of unperturbed reference states.
We applied this approach for the first time to study SRG-evolved chiral  2N and 3N interactions on the isotopic chains of nitrogen, oxygen and fluorine. We find that chiral 3NF at N$^2$LO are crucial in predicting the binding energies of these isotopes and that  the findings of Ref.~\cite{Otsuka2010} apply to other isotopic chains, as well as to the dripline of nitrogen. Within the estimated errors due to the many-body techniques and the dependence on the SRG evolutions, we find a remarkable agreement between our calculations and the experimental energies along all three isotopic chains. 

   Recent works~\cite{Idini2012,Soma13a} clearly show that state of the art SCGF methods can be  extended to the corresponding Gorkov formalism for open shells, which is now underway. This would not only allow direct calculations of semi-magic even-even isotopes with analogous quality as above but would also allow extracting a wealth of information on neighbor isotopes reachable by transfer of one or two nucleons.

{\it Acknowledgements}. This work was supported by the United Kingdom Science and Technology Facilities Council (STFC) under Grant ST/J000051/1 and the Natural Sciences and Engineering Research Council of Canada (NSERC) Grant No. 401945-2011. TRIUMF receives funding via a contribution through the National Research Council Canada. 


\bibliography{bisodu}

\end{document}